# Contribution of Bremsstrahlung Emission from Lyα Clouds to the Microwave Background Fluctuations

Abraham Loeb

Astronomy Department, Harvard University

60 Garden St., Cambridge MA 02138

## ABSTRACT

I calculate the contribution of Bremsstrahlung emission from Lyα absorption clouds to the brightness of the microwave sky. The calculation is based only on the assumption that the clouds below the Lyman-limit are in photoionization equilibrium with a UV background radiation, and avoids any uncertainty about the clumpiness of the gas. I predict a minimum fluctuation amplitude in the Rayleigh-Jeans regime of $\Delta T/T = 10^{-5.5\pm 0.4} \langle J_{21}^2 \rangle^{1/2} (\lambda/5{\rm cm})^2$, which varies over characteristic angular scales $\sim 1\text{-}100''$, where $\lambda$ is the observed wavelength and $\langle J_{21}^2 \rangle^{1/2}$ is a weighted redshift average of the UV background intensity at the Lyman-limit in units of $10^{-21}$ erg cm$^{-2}$ s$^{-1}$ Hz$^{-1}$ sr$^{-1}$. Detection of this signal can be used to map the intergalactic hydrogen distribution and to calibrate the UV background at high redshifts. Existing VLA observations constrain $\langle J_{21}^2 \rangle^{1/2} \lesssim 10^{1.3\pm 0.4}$, unless some of the extended flat-spectrum sources which were detected are Lyα absorption systems.

## 1. Introduction

The Lyα absorption lines in the spectra of QSOs reveal the existence of neutral hydrogen clouds along any line-of-sight throughout the universe (see, e.g. reviews in Blades, Turnshek & Norman 1988). Close QSO pairs and multiple images of gravitationally lensed QSOs show similar absorption spectra, and thus imply that the sizes of the Lyα clouds are $\gtrsim$ 100kpc (Bechtold et al. 1994; Dinshaw et al. 1994, 1995; Fang et al. 1995). When this constraint is combined with the measured HI column density and velocity width of the Lyα absorption lines it follows that the clouds are most likely photoionized by a UV background radiation. Despite its fundamental role in controlling the physical properties of Lyα clouds, the intensity of the UV background is still uncertain to within an order of magnitude. So far, the only empirical method to calibrate this intensity has been the proximity effect, where one uses the fact that the HI abundance in Lyα clouds is diluted near a QSO out to a distance where the background flux starts to dominate over the QSO flux (e.g., Bajtlik, Duncan, & Ostriker 1988; Lu, Wolfe, & Turnshek 1991; Bechtold 1994). Estimates based on this effect imply a UV background intensity at the Lyman-limit of $J_\nu = 10^{-21\pm1}$ erg cm$^{-2}$ s$^{-1}$ Hz$^{-1}$ sr$^{-1}$ (Bechtold 1994); however these estimates may suffer from considerable systematic effects due to velocity shifts in QSO redshifts (Espey 1993), clustering of Lyα clouds (Loeb & Eisenstein 1995), or QSO variability on timescales $\lesssim 10^4$yr (Bajtlik et al. 1988).

In this *Letter* I show that the UV background at high redshifts can be calibrated by measuring the Bremsstrahlung emission from Lyα clouds at microwave frequencies. In §2 I relate this emission to the measured column density distribution of the Lyα forest in a model independent way. The derivation is only based on the assumption that Lyα clouds below the Lyman-limit are in photoionization equilibrium with the UV background. §3 summarizes the implications of this work.

## 2. Bremsstrahlung Emission from Lyα Clouds

In this *Letter* I derive a conservative lower limit to the thermal Bremsstrahlung emission from Lyα absorption clouds. This derivation rests on a single assumption, namely that the clouds are in photoionization equilibrium with a UV radiation background. This assumption is justified by the fact that for the systems under consideration the photoionization time, $\sim 10^4$ yr, is much shorter than the Hubble time. The existence of additional ionization processes due to collisional ionization or photoionization by sources embedded within the clouds, can only increase our estimate for the Bremsstrahlung emissivity. In order to allow the entire cloud to be exposed to the same UV background flux, we consider only Lyα clouds which are optically-thin, i.e. have column densities of neutral hydrogen below the Lyman-limit, $N \leq 10^{17}$ cm$^{-2}$. In addition, we limit our attention only to the redshift interval

$0 \leq z \leq 5$ which was probed so far by observed QSOs. Again, this constraint can only result in an underestimate of the predicted Bremsstrahlung signal.

Because of the particularly high sensitivity of radio detectors, we focus our discussion on the emission of microwave photons. The energy of these photons is much lower than the estimated Ly$\alpha$ cloud temperatures $\sim$ 1-5 eV (Peebles 1993; Miralda-Escudé et al. 1995), and so the Bremsstrahlung spectrum is flat in this regime. The emissivity at any point along the line-of-sight is given by

$$j(\nu) = 1.4 \times 10^{-40} \frac{n_e^2}{(T/5 \times 10^4 \text{K})^{0.5}} \quad \text{erg cm}^{-3} \text{ s}^{-1} \text{ Hz}^{-1} \text{ sr}^{-1}, \qquad (1)$$

where $n_e$ is the number density of free electrons in cm$^{-3}$ and $T$ is the gas temperature. In the range of temperatures and photon frequencies of interest, the Gaunt factor is $\bar{g}_{ff} \approx 6$, with only a logarithmic dependence on these parameters (Karzas & Latter 1961; Rybicki & Lightman 1979). The line-of-sight integral of the emissivity times the volume dilution factor due to the expansion of the universe gives the total surface brightness observed today,

$$J_{\text{Ly}\alpha}(\nu) = \int \frac{j([1+z]\nu)}{(1+z)^3} c dt. \qquad (2)$$

For a hydrogen plasma in photoionization equilibrium we can relate the combination of variables on the right-hand-side of equation 1 to the neutral hydrogen density (Spitzer 1978; Peebles 1993),

$$\frac{n_e^2}{(T/5 \times 10^4 \text{K})^{0.5}} \approx 30 \times J_{21} \times n_{\text{H}}, \qquad (3)$$

where $n_{\text{H}}$ is the HI density in cm$^{-3}$, and $J_{21}(z)$ is the UV background intensity at the Lyman-limit, $J_\nu(z)$, in units of $10^{-21}$ erg cm$^{-2}$ s$^{-1}$ Hz$^{-1}$ sr$^{-1}$. We ignore the additional weak dependence of the full recombination coefficient on temperature ($\propto T^{-0.2}$, cf. Spitzer 1978). By substituting equation 3 in equations 1 and 2 we get

$$J_{\text{Ly}\alpha}(\nu) = 4 \times 10^{-39} \int J_{21} \times \frac{d\Sigma_{\text{HI}}}{dz} \times \frac{dz}{(1+z)^3} \quad \text{erg cm}^{-2} \text{ s}^{-1} \text{ Hz}^{-1} \text{ sr}^{-1}. \qquad (4)$$

where $d\Sigma_{\text{HI}} \equiv n_{\text{H}} c dt$ is the HI column density in the path length interval $cdt$ along the line-of-sight, in units of cm$^{-2}$. Note that this relation avoids any uncertainty about the clumpiness of the gas, as both the recombination rate and the Bremsstrahlung emission rate scale as $n_e^2$. We therefore find that the integrated Bremsstrahlung emission can be expressed in terms of a quantity which is directly inferred from absorption spectra of QSOs, namely the HI column density distribution.

The probability of intersecting a Ly$\alpha$ cloud with an HI column density between $N$ and $N + dN$ in a redshift interval $dz$ around $z \sim 3$ is well fitted by a power-law (Carswell et al.

1984; Tytler 1987),

$$\left(\frac{dP}{dNdz}\right)_{z=3} = AN^{-\beta}, \qquad (5)$$

with $A = 10^{9.1\pm0.3}$ cm$^{[2\beta+2]}$ and $\beta = 1.51 \pm 0.02$ (Sargent, Steidel, & Boksenberg 1989). At this redshift,

$$\left(\frac{d\Sigma_{\rm HI}}{dz}\right)_{z=3} = \int_0^{N_{\rm max}} N\left(\frac{dP}{dNdz}\right)_{z=3} dN = \frac{A}{(2-\beta)} \times N_{\rm max}^{[2-\beta]} = 10^{17.7\pm0.3}{\rm cm}^{-2}, \qquad (6)$$

where we consider systems below the Lyman-limit with $N_{\rm max} = 10^{17}$cm$^{-2}$. Note that systems with $N > N_{\rm max}$ emit at least as much as systems with $N \sim N_{\rm max}$ from their outer, mean-free-path thick, surface which is exposed to the UV background radiation. For spherical clouds, this additional contribution multiplies equation 6 by a correction factor of $(\beta - 1)^{-1} \approx 2$. However, for sheets or filaments that are elongated along the line-of-sight or for systems that contain intrinsic UV sources, the correction factor is bigger. We ignore these factors here. Systems with column densities above the Lyman-limit are likely to be associated with galaxies (e.g., Steidel et al. 1994, 1995; Katz et al. 1995), and so they should occupy much smaller angular scales than low column-density systems.

We next perform the integral in equation 4 over the redshift range $0 \le z \le 5$. According to equation 6, the emission is dominated by the systems close to the Lyman-limit. For these systems we adopt the redshift dependence, $d\Sigma_{\rm HI}/dz \propto (1+z)^\gamma$, with $\gamma = 1.5 \pm 0.4$ (Stengler-Larrea et al. 1995; see also Storrie-Lombardi et al. 1994) when extrapolating equation 6 to other redshifts. We then get from equations 4 and 6 the integrated microwave brightness,

$$J_{\rm Ly\alpha} = 10^{-21.5\pm0.4} \times \langle J_{21} \rangle \quad {\rm erg\ cm}^{-2}\ {\rm s}^{-1}\ {\rm Hz}^{-1}\ {\rm sr}^{-1}, \qquad (7)$$

where

$$\langle J_{21} \rangle \equiv \frac{\int_0^5 J_{21}(z) \times (1+z)^{-[3-\gamma]} dz}{\int_0^5 (1+z)^{-[3-\gamma]} dz}. \qquad (8)$$

The angular brackets in equation 8 can be regarded as a weighted average of the Lyman-limit intensity $J_{21}(z)$ over its redshift evolution.

The brightness in equation 7 is somewhat lower than its potential upper limit based on energy conservation. If the Ly$\alpha$ clouds obtain their thermal energy through photoionization heating, then in a steady state they cannot emit through Bremsstrahlung more thermal energy than they absorb. Since there is roughly one Lyman-limit system per unit redshift, the UV background flux is reprocessed to lower photon energies every unit redshift. As the Bremsstrahlung spectrum is flat up to photon energies $\sim kT \sim 0.5$Ry, the emitted microwave brightness at $z \lesssim 1$ is roughly bounded by the Lyman-limit brightness, $J_\nu = 10^{-21} \times J_{21}$ erg cm$^{-2}$ s$^{-1}$ Hz$^{-1}$ sr$^{-1}$.

The ratio between the microwave brightness of Ly$\alpha$ clouds in equation 7 and the Rayleigh-Jeans brightness of the cosmic background radiation with $T_{\rm CBR} = 2.73$ K is given by,

$$\frac{J_{\rm Ly\alpha}}{(2kT_{\rm CBR}/\lambda^2)} = 10^{-5\pm 0.4} \times \left(\frac{\lambda}{5{\rm cm}}\right)^2 \times \langle J_{21}\rangle, \qquad (9)$$

where $\lambda$ is the observed wavelength. This spectral distortion is well below the COBE limit (Mather et al. 1994), but it should vary on the sky over the characteristic angular scale of Ly$\alpha$ clouds. Despite the lack of an appropriate model for the distribution of cloud sizes, we can still evaluate the *rms* level of fluctuations around the mean brightness, $J_{\rm Ly\alpha}$, due the finite number of clouds along a given line-of-sight. For this purpose, we assume that the clouds are randomly distributed in redshift and column density as indicated by observations (Sargent et al. 1980; Webb & Barcons 1991). In each infinitesimal $(\Delta N, \Delta z)$ bin there is either one or no clouds. The probability of having a cloud with an emission weight $W \propto N \times J_{21}(z)/(1+z)^3$, is $p = (dP/dNdz)\Delta N \Delta z \ll 1$. Therefore, the variance in each bin is $p(1-p)W^2 \approx pW^2$. The integrated variance is then obtained by summing in quadrature the emission from all infinitesimal bins,

$$\langle \Delta J_{\rm Ly\alpha}^2\rangle = \int_0^5 dz \int_0^{N_{\rm max}} dN \frac{dP}{dNdz}\left[4\times 10^{-39} N \times \frac{J_{21}(z)}{(1+z)^3}\right]^2. \qquad (10)$$

This gives a fluctuation amplitude,

$$\frac{\Delta T}{T_{\rm CBR}} \equiv \frac{\langle \Delta J_{\rm Ly\alpha}^2\rangle^{1/2}}{(2kT_{\rm CBR}/\lambda^2)} = 10^{-5.5\pm 0.4} \times \left(\frac{\lambda}{5{\rm cm}}\right)^2 \times \langle J_{21}^2\rangle^{1/2} \qquad (11)$$

where

$$\langle J_{21}^2\rangle \equiv \frac{\int_0^5 J_{21}^2(z) \times (1+z)^{-[6-\gamma]} dz}{\int_0^5 (1+z)^{-[6-\gamma]} dz}. \qquad (12)$$

The fluctuation amplitude $\Delta T/T_{\rm CBR}$ in equation 11 would appear on the characteristic angular scale of Ly$\alpha$ clouds. For clouds with a physical dimension $\sim 100\ell_{100}\ h^{-1}$ kpc (Bechtold et al. 1994; Dinshaw et al. 1994, 1995; Fang et al. 1995; Katz et al. 1995), this scale is $\sim 7'' \times \ell_{100}$ at $z \sim 1$. At low column densities ($\Delta T/T \propto N_{\rm max}^{0.75}$), the Ly$\alpha$ absorption may well be associated with extended large-scale structures that could introduce filamentary fluctuations on scales as large as 1-10 arcminutes (e.g., see images in Miralda-Escudé et al. 1995). In the future, it would be useful to calculate the entire power-spectrum of microwave fluctuations on different angular scales using numerical hydrodynamics simulations of Ly$\alpha$ absorption systems (Cen et al. 1994; Hernquist et al. 1995; Zhang et al. 1995; Miralda-Escudé et al. 1995).

Fomalont et al. (1993) placed tight upper limits on the microwave background anisotropy on 10-100$''$ scales using VLA observations at $\lambda = 3.55$ cm (8.44 GHz). From their limit on Gaussian fluctuations with a coherence angle between 22-60″, $\Delta T/T_{\rm CBR} < 3.0\times 10^{-5}$, we obtain the constraint $\langle J_{21}^2\rangle^{1/2} \lesssim 10^{1.3\pm 0.4}$. This upper limit is well above the expected UV background intensity at low redshifts (Kulkarni & Fall 1993), but is closer to its expected value at high redshifts, $J_{21}(1.6 < z < 4.1) = 10^{0.5\pm 0.5}$ (Bechtold 1994).

The galactic free-free emission at high-latitudes was measured by the COBE satellite on angular scales $\gtrsim 7°$ (Kogut et al. 1995). However, the galactic fluctuations on 1-100″ are currently unknown. Theoretical modeling of these fluctuations (Tegmark & Efstathiou 1995) yields an amplitude on $\sim$ 1-10″ which is smaller than $(\Delta T/T_{\rm CBR})/\langle J_{21}^2\rangle^{1/2}$ in equation 11. It may also be possible to isolate the galactic free-free component based on its correlation with the dust emission (Kogut et al. 1995). Moreover, the Ly$\alpha$ cloud signal around the direction of QSOs should correlate through equation 4 with the column density distribution, $d\Sigma_{\rm HI}(z)/dz$, observed in the spectrum of each QSO. The dominant contamination to the Ly$\alpha$ cloud signal comes, however, from radio sources. The radio source fluctuation amplitude can be as high as $(0.5\text{-}10)\times 10^{-5}$ on $\sim 10''$ scales for $\lambda = $ 2-5cm (Franceschini et al. 1989). This contamination can be partially removed through high-resolution multiple-frequency observations (Fomalont et al. 1993; Windhorst et al. 1993). The Ly$\alpha$ cloud signal can be separated from steep-spectrum sources or from intrinsic microwave background fluctuations based on its spectrum. Bearing in mind that equation 11 provides only a lower limit to the emission signal, it is in fact possible that some of the extended flat-spectrum sources separated by Windhorst et al. (1993) are Ly$\alpha$ absorption systems.

## 3. Conclusions

The minimum contribution of Bremsstrahlung emission from Ly$\alpha$ clouds to the fluctuations of the microwave sky is given by equation 11. The actual emission signal could be substantially higher than this estimate if in addition to the UV background the clouds were also ionized by sources internal to them (such as massive stars or supernovae), if the contribution from clouds with column densities $\gtrsim 10^{17}$ cm$^{-2}$ is significant due to their own ionizing sources or due to their non-spherical geometry, or if there was emission during an early reionization epoch at redshifts $\gtrsim 5$. Existing VLA observations on $\sim 20''$ scales (Fomalont et al. 1993) already constrain $\langle J_{21}^2\rangle^{1/2} \lesssim 10^{1.3\pm 0.4}$, unless some of the extended flat-spectrum sources which were detected (Windhorst et al. 1993) are Ly$\alpha$ absorption systems. It is therefore possible that future experiments will be able to place more stringent limits on the UV background flux than currently exist (Bechtold 1994). The main challenge of such experiments would be to separate the Ly$\alpha$ cloud signal from the radio source background. Since the anisotropy constraints are tighter on larger angular scales, it is important to apply existing hydrodynamic simulations of Ly$\alpha$ clouds (e.g., Hernquist et al. 1995; Zhang et al. 1995; Miralda-Escudé et al. 1995) to the study of the Bremsstrahlung fluctuations on all angular scales, particularly beyond $1'$ (see observational constraints by Uson & Wilkinson 1984, Readhead et al. 1989, Myers et al. 1993, and Subrahmanyan et al. 1993). Such simulations may also be able to quantify the probability distribution of the fluctuation amplitude on a given angular scale.

The potential detection of Bremsstrahlung emission from Ly$\alpha$ clouds will allow direct mapping of their hydrogen column density distribution. In addition, calibration of the UV background flux $J_{21}$ could shed light on the physical conditions in the clouds, and may also constrain the baryonic density parameter $\Omega_b$ when combined with results from hydrodynamic simulations (see discussions in Hernquist et al. 1995 and Miralda-Escudé et al. 1995). Finally, it would be interesting to cross-correlate the microwave fluctuations of Ly$\alpha$ clouds with maps of faint blue galaxies in order to uncover any physical relation between these two populations (Fang et al. 1995).

I thank D. Eisenstein, G. Field, U. Seljak, and G. Rybicki for useful discussions. This work was supported in part by the NASA ATP grant NAG5-3085.

---